\documentclass[journal,letterpaper,onecolumn]{IEEEtran}

\usepackage{amsmath}
\usepackage{amssymb}
\usepackage{graphicx}
\usepackage{bm}
\usepackage{amsthm}
\usepackage{algorithm}%
\floatname{algorithm}{Protocol}
\usepackage{color}
\usepackage{dsfont}

\newcommand\be{\begin{equation}}
\newcommand\ee{\end{equation}}
\newcommand{\bea}{\begin{eqnarray}}
\newcommand{\eea}{\end{eqnarray}}

\newcommand\eps{\epsilon}
\newcommand{\tr}{\mathop{\rm Tr}}
\newcommand{\Tr}{\mathop{\rm Tr}}

\newcommand{\altqed}{\hfill{\small $\blacksquare$}}

\def\NN{\mathbb{N}}
\def\PP{\mathbb{P}}
\def\EE{\mathbb{E}}

\def\unity{\mathbb{I}}
\def\one{\unity}

\def\be{\begin{equation}}
\def\ee{\end{equation}}

\newtheorem{theorem}{Theorem}[section]
\newtheorem{lemma}{Lemma}[section]

\usepackage{times}
\usepackage{color}

\newcommand{\calM}{{\mathcal{M}}}

\newcommand{\calL}{{\mathcal{L}}}
\newcommand{\calH}{{\mathcal{H}}}
\newcommand{\calK}{{\mathcal{K}}}

\newcommand{\calZ}{{\mathcal{Z}}}
\newcommand{\calS}{{\mathcal{S}}}

\newcommand{\IGamma}{{I_{\Gamma}}}
\newcommand{\PGamma}{{P_{\Gamma}}}
\newcommand{\SizeIGamma}{{|I_{\Gamma}|}}

\begin{document}

\bibliographystyle{IEEETrans}


\title{The strong converse theorem for the product-state capacity of quantum channels with ergodic Markovian memory}

\author{
    \vspace{2mm}
    \IEEEauthorblockN{Tony Dorlas\IEEEauthorrefmark{1} and Ciara Morgan\IEEEauthorrefmark{2}}\\
    \vspace{2mm}
    \IEEEauthorrefmark{1}Dublin Institute for Advanced Studies, 10 Burlington Road, Dublin 4. Ireland.\\
    \IEEEauthorrefmark{2}School of Mathematics and Statistics, University College Dublin, Belfield, Dublin 4. Ireland.
}

\date{\today}
\maketitle

\begin{abstract}
Establishing the strong converse theorem for a communication channel confirms that the capacity of that channel, that is, the maximum achievable rate of reliable information communication, is the ultimate limit of communication over that channel. Indeed, the strong converse theorem for a channel states that coding at a rate above the capacity of the channel results in the convergence of the error to its maximum value $1$ and that there is no trade-off between communication rate and decoding error. Here we prove that the strong converse theorem holds for the product-state capacity of quantum channels with ergodic Markovian correlated memory. 
\end{abstract}


\section{Introduction}

Establishing the optimal communication rate at which information can be reliably transmitted over noisy quantum channels is a question of central importance in quantum information theory. The achievability, or direct part, of a channel coding theorem, establishes a rate of communication below which the decoding error tends to zero in the limit of large block length. This raises the natural question of whether a larger rate can be achieved with a decoding error which is not equal to zero but less than $1$. In other words, the question of whether an error-rate trade-off is possible emerges. The strong converse theorem addresses this question and when established for a particular channel, confirms that there can be no such error-rate trade-off for that channel.

Traditionally, noisy communication is modelled by the repeated application of a particular channel to an encoded message. This type of channel is referred to as memoryless, and this assumption of independent successive channel uses is considered to be unrealistic. 

In this work we establish the strong converse theorem for the communication of classical information encoded into product quantum states and transmitted over a particular class of quantum channels with memory. That is, we are concerned with establishing the strong converse theorem corresponding to the product-state capacity of these channels with memory. The behaviour of the quantum channels, denoted by $\Phi^{(n)}$, is modelled by an ergodic Markov chain, on a finite state space $I$, specified by an invariant distribution $\{ \gamma_i\}_{i \in I}$  and transition matrix $\{ q_{i,i'}\}_{i \in I}$ over a finite set of quantum channels $\{\Phi_{i}\}_{i \in I}$. Indeed this invariant distribution initiates the channel sequence, with the transition matrix $\{ q_{i,i'}\}_{i \in I}$ governing the subsequent behaviour.
Moreover we consider the particular case where the overall behaviour of the channel $\Phi^{(n)}$ is determined by an \emph{ergodic} Markov chain. That is, a Markov chain which is a periodic and irreducible, resulting in the convergence of the $n$-step transition probability $q_{i,i'}^{(n)}$ to equilibrium in the asymptotic limit, i.e.  $q_{i,i'}^{(n)} \rightarrow \gamma_i$, as $n \rightarrow \infty$. In this case the quantum channel $\Phi^{(n)}$ is considered to be \emph{forgetful}. 

To provide some background on this type of quantum channel with memory, we note that Macchiavello and Palma \cite{MP02} were the first to consider a Markovian noise correlation model for communication over quantum channels. Here they study the entanglement-assisted classical capacity for a quantum channel with so-called partial memory, where the channel is written as a convex combination of a sequence of uncorrelated depolarising channels, i.e. a memoryless depolarising channel, and a sequence of correlated depolarising channels. They showed a higher mutual information can be achieved by entangling two successive uses of this channel. 

Bowen and Mancini \cite{BowenMancini04} considered a more general model for noise correlation which includes Markovian noise correlations as a special case. In particular, taking the set of possible channels to be those which can be written as unitary Kraus operators, with error probabilities given by steady-state probabilities for the underlying Markov chain, they recover the HSW (Holevo-Schumacher-Westmoreland) \cite{Holevo98,SW97} capacity bound. In \cite{Markovmem09} Datta and Dorlas generalise this result to arbitrary Markov chains, generalising the HSW capacity. 

Previously, Datta and Dorlas \cite{DD07} had established the coding theorem and weak converse for the product-state capacity of a class of channels with so-called long-term memory, given by a convex combination of memoryless channels. Later Dorlas and Morgan \cite{DM11} also considered another type of channel with long-term memory, namely the periodic channel and showed that for a particular realisation of this channel in terms of amplitude damping channels, the strong converse in fact does not hold. 

On the other hand, the strong converse theorem has been shown to hold for the product-state capacity for all {\em memoryless} quantum channels. This result was proved independently by Winter \cite{Winter99} using the so-called method of types, generalising the technique of Wolfowitz \cite{Wolfowitz} for classical channel coding, and by Ogawa and Nagaoka \cite{ON99} generalising Arimoto's method \cite{Arimoto}. 

Indeed, it is notable that the non-commutative generalisation of Arimoto method and it's connection to R\'enyi entropy and divergence, has also lead to successes in the attempt to establish a strong converse theorem for certain memoryless quantum channels with {\em arbitrary} input states. We point to K\"onig and Wehner \cite{KW09} for the first such proof and for a treatment of the general open problem. We note the following review article \cite{CGLM14} for further details involving communication over quantum channels with memory.

In this article, we follow similar lines to Winter \cite{WinterThesis99} to prove that the strong converse holds for the product-state capacity of channels with ergodic Markovian correlated memory. We note that proving this result using the alternative Arimoto method might also be of interest, as well as establishing a strong converse theorem for this noise model in the case where restrictions on the input states and/or on the type of memory are lifted. In the latter case \emph{quantum} memory, considered by Bowen and Mancini \cite{BowenMancini04} and, more generally by Kretschmann and Werner \cite{KW05}, could be considered.

\section{Definition of the channel and statement of the theorem}

We consider quantum channels with Markovian memory as first introduced by Macchiavello and Palma
\cite{MP02}. The general classical capacity of such channels was derived in \cite{Markovmem09}.

Let there be given a Markov chain on a finite state space $I$ with transition probabilities
$\{q_{ii'}\}_{i,i' \in I}$ and let $\{\gamma_i\}_{i \in I}$ be an invariant distribution for this
chain, i.e.
\begin{equation} \gamma_{i'} = \sum_{i \in I} \gamma_{i} q_{ii'}.
\label{eqdist}
\end{equation}
Moreover, let $\Phi_i: {\cal B}({\cal H}) \to {\cal B}({\cal K})$ be given completely positive
trace-preserving (CPT) maps for each $i \in I$, where $\cal H$ and $\cal K$ are finite-dimensional
Hilbert spaces. We also consider the tensor product algebras ${\cal A}_n = {\cal B}({\cal
H}^{\otimes n})$ and the infinite tensor product C$^*$-algebra obtained as the strong closure
\begin{equation} {\cal A}_\infty = \overline{ \bigcup_{n=1}^\infty
{\cal A}_n}, \end{equation} where we embed ${\cal A}_n$ into ${\cal A}_{n+1}$ in the obvious way.
Similarly, we define ${\cal B}_n = {\cal B}({\cal K}^{\otimes n})$ and ${\cal B}_\infty$. A
{\em{state}} on an algebra ${\cal{A}}$ is a positive linear functional $\phi$ on ${\cal{A}}$ with
$\phi({\unity})= 1$, where $\unity$ denotes the identity operator. If ${\cal{A}}$ is finite-dimensional
then there exists a density matrix $\rho_\phi \in {\cal A}$ such that $\phi(A) = {\tr}(\rho_\phi
A),$ for any $A \in {\cal{A}}$. We denote the states on ${\cal A}_\infty$ by ${\cal S}({\cal
A}_\infty)$, those on ${\cal A}_n$ by ${\cal S}({\cal A}_n)$, etc.

We now define a quantum channel with Markovian-correlated noise by the CPT map $\Phi_\infty: {\cal
S}({\cal A}_\infty) \to {\cal S}({\cal B}_\infty)$ on the states of ${\cal A}_\infty$ by
\begin{equation} (\Phi_\infty) (\phi) (A) = \sum_{i_1,\dots,i_n
\in I} \gamma_{i_1} q_{i_1i_2} \dots q_{i_{n-1}i_n} \tr \left[ (\Phi_{i_1} \otimes \dots \otimes
\Phi_{i_n})(\rho_{\phi_n}) \, A \right] \label{mem}
\end{equation} for $A \in {\cal B}_n$. Here, $\phi_n$ is the
restriction of $\phi$ to ${\cal A}_n$ and $\rho_{\phi_n}$ its density matrix. It is easily seen,
using the property (\ref{eqdist}), that this definition is consistent and defines a CPT map on the
states of ${\cal A}_\infty$, and moreover, that it is translation-invariant (stationary).

We denote the transpose action of the restriction of $\Phi_\infty$ to ${\cal S}({\cal A}_n)$ by
$\Phi^{(n)}: {\cal B}({\cal H}^{\otimes n}) \to {\cal B}({\cal K}^{\otimes n})$, i.e.,
$${\tr}\bigl(\Phi^{(n)}(\rho_\phi)A \bigr) =(\Phi_\infty(\phi))(A),$$
for a density matrix $\rho_\phi \in {\cal{B}}({\cal{H}}^{\otimes n})$, $\phi \in  {\cal S}({\cal
A}_n)$.

Thus
\begin{equation}
\Phi^{(n)}(\rho^{(n)}) = \sum_{i_1,\dots,i_n \in I} \gamma_{i_1} q_{i_1i_2} \dots q_{i_{n-1}i_n}
(\Phi_{i_1} \otimes \dots \otimes \Phi_{i_n})(\rho^{(n)}). \label{mem2}\end{equation}

In the following we assume that the Markov chain $(q_{ii'})$ is irreducible and aperiodic. The
quantum channel is then \textit{forgetful} in the terminology of Kretschmann and Werner
\cite{KW05}. In that case it was proved in \cite{Markovmem09} (and in a more general setting in \cite{KW05})
that the classical capacity of the channel is given by

\begin{equation} \label{capacity}
{\chi}^*(\Phi) = \lim_{n \to \infty} \frac{1}{n} \sup_{\{p_j^{(n)},\rho_j^{(n)}\}}
\chi(\{p_j^{(n)},\Phi^{(n)} (\rho_j^{(n)})\})  \end{equation} where
\begin{equation} \label{Holevoquantity}
\chi(\{p_j^{(n)},\Phi^{(n)}(\rho_j^{(n)})\}) = S \left( \sum_{j=1}^{J(n)} p_j^{(n)}
\Phi^{(n)}(\rho_j^{(n)}) \right) - \sum_{j=1}^{J(n)} p_j^{(n)} S(\Phi^{(n)}(\rho_j^{(n)}))
\end{equation} is the Holevo quantity for a finite chain. Analogously, the product-state capacity is
given by the same expression but where the supremum in (\ref{capacity}) is restricted to product
states $\rho_j^{(n)}$.

Here we prove the strong converse for product states of such a channel, analogous to Winter's
theorem, \cite{WinterThesis99}, Theorem~II.7.

\begin{theorem}[Strong Converse] \label{theorem:StrongConverse}
Let $(f^{(n)},D^{(n)})_{n \in \NN}$ be a sequence of product-state codes given by maps $f^{(n)}:\,
{\cal M}_n \to {\cal S}({\cal H}^{\otimes n})$ and decoding operator maps  $D^{(n)}:\,{\cal M}_n
\to {\cal B}({\cal K}^{\otimes n})$ with $\sum_{w \in {\cal M}_n} D^{(n)}(w) \leq \unity$. Here
${\cal M}_n$ is a collection of codewords for each $n$ containing $N(n) = |{\cal M}_n|$ codewords.
Suppose that the code has error probability less than $\lambda\in(0,1)$. Then, for all $\epsilon>0$
there is $n_0=n_0(\lambda,\epsilon)$ such that for every $n\geq n_0$,
\begin{equation}
\log N(n) \leq n(\chi^*(\Phi) + \epsilon).
\end{equation}
\end{theorem}

\begin{IEEEproof}
Let $(f,D)$ be an $(n, \lambda)$-code, where $f$ is the encoding map  $f: \mathcal{M} \rightarrow
\calL(\calH)^{\otimes{n}}$ with $f(w) = f_1(w) \otimes  \cdots \otimes f_n(w)$ and $D = (D(w))_{w
\in \calM}$, such that the error probability
\[
e(f,D) = \max_{w \in \calM}  \left[1 -  \left( \Phi^{(n)} (f(w))\, D(w) \right) \right] \leq
\lambda.
\]
(We suppress the dependence on $n$.)

Fix $\eps, \delta >0 $ and let $l_0 \in \NN$ be so large that \begin{equation} \label{ergodic}
|q^{(l_0)}_{ij} - \gamma_j| < \delta^3 \gamma_j \end{equation}  for all $i,j \in I$, where we
define, for general $n \in \NN$,
\begin{equation} q^{(n)}_{ij} = \sum_{i_2,\dots,i_{n-1} \in I} \gamma_{ii_2} \gamma_{i_2 i_3}
\dots q_{i_{n-1}j}. \end{equation}

Then let $n_0 \gg l_0$ be large enough so that \begin{equation} \chi^{(n_0)}< n_0 (\chi^*(\Phi) + \eps),
\end{equation} where $\chi^{(n_0)}$ is the supremum (over product state ensembles) on the right-hand side
of (\ref{capacity}).

Given $\theta > 0$, consider a $\theta$-fine partition $\calZ = (\calZ_j)_{j=1}^J$ of the compact
space $\calS(\calK^{\otimes n_0})$, and fix $\tilde{\sigma}_j^{(n_0)} \in \calZ_j$ for each $j \in
\{1,\dots,J\}$. Consider the states \begin{equation} \Phi^{(n_0)}(\rho_k^{(n_0)}(w)) \in \calS(\calK^{\otimes n_0}),
\end{equation} where $k=1,\dots,m$ with $m = [n/(n_0+l_0)]$ and where
\begin{equation}\label{defrho}  \rho_{k,l}(w) = f_{(n_0 + l_0)(k-1)+l}(w) \mbox{ and } \rho_k^{(n_0)}(w) =
\rho_{k,1}(w) \otimes \dots \otimes \rho_{k,n_0}(w). \end{equation}

Given a \lq class' $\Gamma$, that is,  a subset of $\{ 1, \cdots, J\}$, let \\ $\IGamma = \{k :
\Phi^{(n_0)}(\rho_k^{(n_0)}(w)) \mathrm{\; has \; class\; }
\Gamma \}$, that is
\begin{equation}
\IGamma = \{k \leq m: \{ j: \exists w \in \calM: \Phi^{(n_0)}(\rho_k^{(n_0)}(w)) \in \calZ_j  \} = \Gamma \}.
\end{equation}

For each $\Gamma$, we define the \textbf{type} of $f(w)$ on $\Gamma$ to be the probability measure $P$ on
$\{1,\dots,J\}$ such that for all $j \in \{1,\dots,J\}$,
\begin{equation} \# \{k \in \Gamma:\, \Phi^{(n_0)}(\rho_k^{(n_0)}(w)) \in \calZ_j\} = \SizeIGamma \,P(j).
\end{equation}

Consider the $\calZ$-types of the product states $\bigotimes_{k \in \IGamma}
\Phi^{(n_0)}(\rho_k^{(n_0)}(w))$ over the positions in $\IGamma$.
The number of types of these states is bounded by $(\SizeIGamma + 1)^J$. For each $\IGamma \neq
\emptyset$, we can therefore select a type $\PGamma$ which is realised for at least $(\SizeIGamma
+1)^{-J} |\calM|$ of the codewords, and reduce the code to these codewords. The resulting code
$\calM'$ then has a unique type $\PGamma$ for each $\Gamma$, and the following bound on the number
of codewords holds
\begin{equation}\label{eq:RatioCodeSize}
|\calM'| \geq \prod_{\gamma =1}^{2^J} (\SizeIGamma +1)^{-J} |\calM| \geq (m+1)^{-J2^J} |\calM|.
\end{equation}

For each $k \in \IGamma$ we choose a state $\sigma_{k,j}^{(n_0)} = \Phi^{(n_0)}(\rho_k^{(n_0)}(w)) \in \calZ_j$ and define
\begin{equation}
\bar{\sigma}_{k, \Gamma}^{(n_0)} = \sum_j \PGamma(j) \sigma_{k,j}^{(n_0)}.
\end{equation}
We also put
\begin{equation}
\tilde{\sigma}_{\Gamma}^{(n_0)} = \sum_j \PGamma(j) \tilde{\sigma}_j^{(n_0)},
\end{equation}
where $\tilde{\sigma}_j^{(n_0)}$ is the reference state in $\calZ_j$ chosen above.

Classes $\Gamma$ are considered \emph{large} if
\begin{equation}
\SizeIGamma \geq m 2^{-J} \eps.
\end{equation}
For \emph{large} classes $\Gamma$ we define, given $\delta > 0$, the typical projection
$\tilde{\Pi}_{\Gamma,\delta}$ by
\begin{equation} \label{deftypical}
\tilde{\Pi}_{\Gamma,\delta} = \Pi_{\tilde{\sigma}_\Gamma^{(n_0)},\delta} = \sum_{(p_k)_{k \in
\IGamma} \in \tilde{\cal T}_{\Gamma,\delta}} \,\bigotimes_{k \in \IGamma} \tilde{\pi}_{\Gamma,p_k},
\end{equation}
where $\tilde{\sigma}_\Gamma^{(n_0)} = \sum_{p=1}^{d^{n_0}} \lambda_{\Gamma,p} \,\tilde{\pi}_{\Gamma,p}$ is the
diagonalisation of $\tilde{\sigma}_\Gamma^{(n_0)}$, and the typical set $\tilde{\cal
T}_{\Gamma,\delta}$ is given by \begin{equation} \label{typicalset}  \tilde{\cal T}_{\Gamma,\delta}
= \left\{ (p_k)_{k \in \IGamma}:\, \big|\#\{k \in \IGamma:\,p_k = p\} - \SizeIGamma \lambda_{\Gamma,p} \big| \leq
\SizeIGamma \delta \right\}. \end{equation}

Then,  by Lemma~\ref{lemma:TypicalSpace}, \begin{equation} 2^{-\SizeIGamma (S(\tilde{\sigma}_\Gamma) +
\eps)} \leq \tilde{\Pi}_{\Gamma,\delta} \, \left(\tilde{\sigma}_\Gamma^{(n_0)} \right)^{\otimes \SizeIGamma}
\, \tilde{\Pi}_{\Gamma,\delta} \leq 2^{-\SizeIGamma (S(\tilde{\sigma}_\Gamma) - \eps)}.
\end{equation}

Now we can write
\begin{equation} \label{expansion}
\begin{split} \lefteqn{
\tr \left( \Phi^{(n)}(f^{(n)}(w)) \tilde{\Pi}_{\Gamma,\delta} \right) =} \\ &= \sum_{i_1,\dots,i_{m+1}}
\gamma_{i_1} \tr \left( \sigma_1^{(n_0)}(i_1,i_2) \otimes \dots \otimes
\sigma_m^{(n_0)}(i_{m},i_{m+1})\,\tilde{\Pi}_{\Gamma,\delta} \right),
\end{split}
\end{equation} where
\begin{eqnarray} \label{defsigmaIJ} \sigma_k^{(n_0)}(i,i') &=& \sum_{i_2,\dots,i_{n_0}}  q_{i,i_2} q_{i_2 i_3} \dots
q_{i_{n_0-1},i_{n_0}} q^{(l_0)}_{i_{n_0},i'}
\\ && \qquad \times \Phi_{i}(\rho_{k,1}(w)) \otimes \dots \otimes \Phi_{i_{n_0}}(\rho_{k,n_0}(w)).
\end{eqnarray}

Notice that \begin{eqnarray*} \sigma_k^{(n_0)}(i) &=&  \sum_{i' \in I}  \sigma_k^{(n_0)}(i,i') \\
&=& \sum_{i_2,\dots,i_{n_0}} q_{i,i_2} q_{i_2 i_3} \dots q_{i_{n_0-1},i_{n_0}}
\\ && \qquad \times \Phi_{i}(\rho_{k,1}(w)) \otimes \dots \otimes \Phi_{i_{n_0}}(\rho_{k,n_0}(w))
\end{eqnarray*} is a state since the trace equals
$$  \sum_{i_2,\dots,i_{n_0}} q_{i,i_2} q_{i_2 i_3} \dots q_{i_{n_0-1},i_{n_0}} = 1. $$
Moreover,
$$ \sum_{i \in I} \gamma_i \tr (\sigma_k^{(n_0)} (i,i')) = \sum_{i_1,\dots,i_{n_0}} \gamma_{i_1}
q_{i_1,i_2} q_{i_2 i_3} \dots q_{i_{n_0-1},i_{n_0}} q^{(l_0)}_{i_{n_0} i'} = \gamma_{i'}. $$ Also,
the condition $$ \sigma^{(n_0)}_k(i,i') \leq (1+\delta^3) \gamma_{i'} \sigma^{(n_0)}_k(i)  $$
follows from (\ref{ergodic}). It therefore follows from Lemma~\ref{lemma:MarkovWeakLaw} that
$$ \tr \left( \Phi^{(n)}(f^{(n)}(w)) \, (\tilde{\Pi}_{\Gamma,\delta + \theta} \otimes \one ) \right) \geq 1-2 \delta. $$

We now define \begin{equation} \bar{\Pi}_\delta = \bigotimes_{\Gamma\, {\rm large}}
(\tilde{\Pi}_{\Gamma, \delta} \otimes \one^{(l_0)}) \, \bigotimes_{\Gamma\,{\rm small}} \one^{(n_0
+ l_0)} \otimes \one^{(n-m(n_0+l_0)}. \end{equation} It then follows from
$$ \bigotimes_{\Gamma}\, ( \one - \Pi_\Gamma) \geq \one - \sum_\Gamma\, (\Pi_\Gamma \otimes \one_{\IGamma^c}, $$
that
\begin{equation} \label{ineq} \Tr \left( \Phi^{(n)}(f(w)) \,\bar{\Pi}_{\delta + \theta} \right) > 1- 2^{J+1} \delta.
\end{equation}

By Lemma~\ref{lemma:TypicalSpace},
\begin{equation} \tr \tilde{\Pi}_{\Gamma,\delta + \theta} \leq 2^{\SizeIGamma
(S(\tilde{\sigma}_\Gamma^{(n_0)}) + \eps)} \end{equation} if $\SizeIGamma$ is large enough. Since
$$ \left\| \tilde{\sigma}_\Gamma^{(n_0)} - \bar{\sigma}_{k,\Gamma}^{(n_0)} \right\|_1 < \theta $$
it follows from Lemma~\ref{lemma:Continuity} that
\begin{equation} \tr \tilde{\Pi}_{\Gamma,\delta + \theta} \leq \prod_{k \in \IGamma} 2^{
(S(\bar{\sigma}_{k,\Gamma}^{(n_0)}) + 2\eps} \end{equation} if $\SizeIGamma$ is large and $d^{n_0}
\eta(\theta/d^{n_0}) < \eps$. Taking products we have
\begin{equation} \label{tracePibound} \tr \bar{\Pi}_{\delta + \theta} \leq \prod_{\Gamma\, {\rm large}} \prod_{k \in \IGamma}
2^{\SizeIGamma (S(\bar{\sigma}_{k,\Gamma}^{(n_0)}) + 2\eps)}  d^{n - m n_0 + n \eps}.
\end{equation}

Now consider the code $(f',D')$, where $f' = f_{|_{\calM'}}$ and where $$ D'(w) =
\bar{\Pi}_{\delta+\theta} D(w) \bar{\Pi}_{\delta + \theta} \mbox{ for } w \in \calM'. $$

By Lemma~\ref{lemma:TenderOperator} and (\ref{ineq}),
$$ \left\| \Phi^{(n)}(f(w)) - \bar{\Pi}_{\delta + \theta} \left( \Phi^{(n)}(f(w)) \bar{\Pi}_{\delta +
\theta} \right) \right\|_1 \leq 4 \sqrt{2^J \delta}. $$ Therefore
\begin{equation}
\begin{split}
e(f', D') &=  \max_{w \in \calM'}  \left[1 -  \tr \left( \Phi^{(n)}(f(w)) D'_m \right) \right] \\
& \leq \max_{w \in \calM} \left[ 1 - \tr (\Phi^{(n)}(f(w))) D_m) \right]  + 4 \sqrt{2^J \delta} \\
&\leq \lambda + 4 \sqrt{2^{J} \delta} = \lambda'.
\end{split}
\end{equation}
Clearly, if $\delta$ is small enough, $\lambda' < 1$.

It follows from Lemma~\ref{lemma:EntropyTypical} that
\begin{equation} \tr \left( f(w) \, \Pi_{{\rm ent}, \delta} \right) > 1-\delta. \end{equation}
We also have
\begin{equation}
\Pi_{{\rm ent},\delta}\,\Phi^{(n)}(f(w)) \Pi_{{\rm ent},\delta}
         \leq \prod_{k=1}^m 2^{-S(\Phi^{(n_0)}(\rho_k^{(n_0)})) + 2\delta} \,\Pi_{{\rm ent},\delta}.
\end{equation}

By the Shadow bound (Lemma~\ref{lemma:ShadowBound}), it follows that
\begin{equation}
\begin{split}
\tr D'(w) &\geq (\lambda' - 4 \sqrt{\delta}) \prod_{k=1}^m 2^{S(\Phi^{(n_0)}(\rho_k^{(n_0)})) - 2\delta}
\\ & \geq \prod_{k=1}^m 2^{S(\Phi^{(n_0)}(\rho_k^{(n_0)})) - \eps} \\
&\geq \prod_{\Gamma} \prod_{k \in \Gamma} 2^{\sum_{j=1}^J P_\Gamma(j) S(\sigma_{k,j}^{(n_0)}) - 2 \eps}
\end{split} \end{equation}
for  $m$ large enough and $d^{n_0} \eta(\theta/d^{n_0}) , \eps$.

Combining this with (\ref{tracePibound}) we have
\begin{equation}
\begin{split}
|\calM' | &\leq \prod_{\Gamma} \prod_{k \in \IGamma} 2^{-\sum_{j=1}^J P_\Gamma(j)
S(\sigma_{k,j}^{(n_0)}) + 2 \eps} \tr \sum_{w \in \calM'} D'(w) \\
&\leq \prod_{\Gamma} \prod_{k \in \IGamma} 2^{-\sum_{j=1}^J P_\Gamma(j)
S(\sigma_{k,j}^{(n_0)}) + 2 \eps} \tr \bar{\Pi}_{\delta+\theta} \\
&\leq \prod_{\Gamma} \prod_{k \in \IGamma} 2^{-\sum_{j=1}^J P_\Gamma(j) S(\sigma_{k,j}^{(n_0)}) + 2
\eps} \prod_{\Gamma\, {\rm large}} \prod_{k \in \IGamma} 2^{S(\bar{\sigma}_{k,\Gamma}^{(n_0)}) +
2\eps}  d^{n - m n_0 + n \eps}
\\ &\leq \prod_{\Gamma\, {\rm large}} \prod_{k \in \IGamma} 2^{
S(\bar{\sigma}_{k,\Gamma}^{(n_0)}) - \sum_{j=1}^J P_\Gamma(j) S(\sigma_{k,j}^{(n_0)}) + 4 \eps}
d^{n - m n_0 + n \eps}.
\end{split}
\end{equation}

This yields
$$ |\calM' | \leq 2^{n (\chi^*(\Phi) + (4 + 5 \log d) \eps)} $$
since
$$ \prod_{k \in \IGamma} 2^{
S(\bar{\sigma}_{k,\Gamma}^{(n_0)}) - \sum_{j=1}^J P_\Gamma(j) S(\sigma_{k,j}^{(n_0)})} \leq
2^{\SizeIGamma \chi^{(n_0)}} $$ and
$$ d^{n-m n_0 + n\eps} < \prod_{\Gamma {\rm large}} 2^{5 \eps \log d} $$ since
$$ \sum_{\Gamma {\rm large}} \SizeIGamma > (1-\eps) n $$
and we can take
$$ \frac{3 \eps}{1-\eps} + \frac{n- m n_0}{(1-\eps)n} < 5 \eps $$
by taking $m$ large and $n_0 \gg l_0$.


Finally, using (\ref{eq:RatioCodeSize}) we have
\begin{equation}
|\calM | \leq (m+1)^{J2^J} |\cal M'|
\end{equation}
so that for $m$ large enough
\begin{equation}
|\calM | \leq 2^{n (\chi^*(\Phi) + 5(1 + \log d) \eps)}.
\end{equation}

\end{IEEEproof}

\subsection{Lemmata}

\begin{lemma}\label{lemma:TypicalSpace}
For every state $\rho$ and $\delta, \eps >0$ if $n$ is large enough,
\begin{equation}
  \tr \Pi^n_{\rho,\delta} \leq 2^{n(S(\rho)+\eps)}.
\end{equation}
\end{lemma}

\begin{lemma}[Generalised weak law]
  \label{lemma:MarkovWeakLaw}
Let $\sigma_k(i,i')$, where $k=1,\dots,m$ and $i,i' \in I$ be positive operators on a
finite-dimensional Hilbert space $\calH$ ($\dim(\calH) = d$) such that
\begin{equation} \label{consistency}
\sum_{i \in I} \gamma_i \tr(\sigma_k(i,i')) = \gamma_{i'} \end{equation}  for all $k$ and all $i'
\in I$, and such that for all $k$ and $i \in I$, $\sigma_k(i)$ defined by \begin{equation}
\label{sigmastate} \sigma_k(i) = \sum_{i' \in I} \sigma_k(i,i') \end{equation}  is a state.
Moreover, assume that
\begin{equation} \label{sigmabound} (1-\delta^3) \gamma_{i'} \sigma_k(i) \leq \sigma_k(i,i') \leq (1+\delta^3)
\gamma_{i'} \sigma_k(i) \end{equation} for all $k$ and all $i,i' \in I$.  Define the states
$\sigma^{(m)}$ on $\calH^{\otimes m}$ by
\begin{equation} \label{defsigma_m}
\sigma^{(m)} = \sum_{i_1, \dots, i_m, i_{m+1} \in I} \gamma_{i_1} \sigma_1(i_1,i_2) \otimes \dots
\otimes \sigma_m(i_m,i_{m+1}). \end{equation}

Given a subset $S \subset \{1,\dots,m\}$, set
$$ \bar{\sigma} = \sum_{i \in I} \frac{1}{|S|} \sum_{k \in S} \gamma_i \sigma_k(i) $$
and suppose that $$ || \bar{\sigma} - \tilde{\sigma} ||_\infty < \theta. $$ Diagonalising
$\tilde{\sigma} = \sum_{p=1}^{d} \tilde{q}_p\,\pi_p$, define the typical projection
\begin{equation} 
\Pi_{S,\delta} = \sum_{(p_k)_{k \in S} \in \tilde{\cal T}_{S,\delta}} \left(\bigotimes_{k \in S} \pi_{p_k} \right)
\otimes \one_{S^c}, \end{equation}  where
$$ \tilde{\cal T}_{S,\delta} = \left\{(p_k)_{k \in S}:\, \big| \#\{k \in S:\,p_k = p\} - |S| \tilde{q}_p \big| \leq |S|
\delta \right\}. $$ Then, if $|S|$ is large enough,
$$ \tr\left( \sigma^{(m)} \Pi_{S,\delta + \theta} \right) > 1-2\delta. $$
\end{lemma}

\begin{IEEEproof} For a state $\rho$ on $\calH$ let $\kappa(\rho) = \sum_{p=1}^d \pi_p\, \rho \,\pi_p$ and let
$\kappa_S = \kappa^{\otimes |S|}$, i.e. for a state $\rho^{(m)}$ on $\calH^{\otimes m}$,
$$ \kappa(\rho^{(m)}) = \sum_{(p_k)_{k \in S}} \left( \bigotimes_{k \in S} \pi_{p_k} \otimes \one_{S^c} \right)
\rho^{(m)} \left( \bigotimes_{k \in S} \pi_{p_k} \otimes \one_{S^c} \right). $$ Put
$$\tilde{\sigma}^{(m)} = \kappa_S(\sigma^{(m)}). $$
Then, since $\kappa_S(\Pi_{S,\delta}) = \Pi_{S,\delta}$, $ \tr(\sigma^{(m)} \Pi_{S,\delta}) = \tr
(\tilde{\sigma}^{(m)} \Pi_{S,\delta}), $ and
$$ \tilde{\sigma}^{(m)} = \sum_{i_1, \dots, i_m, i_{m+1} \in I}
\gamma_i \tilde{\sigma}_1(i_1,i_2)  \otimes \dots \otimes \tilde{\sigma}_m(i_m,i_{m+1}),
$$
where $$ \tilde{\sigma}_k(i,i') = \kappa(\sigma_k(i,i')) = \sum_{p=1}^d \lambda_k(i,i'; p) \pi_p
$$ is diagonal w.r.t. the $\tilde{\sigma}$ basis for $k \in S$, and where  $\tilde{\sigma}_k(i,i') =
\sigma_k(i,i')$ for $k \notin S$. Moreover,
$$ q_p = \sum_{i,i' \in I} \frac{1}{|S|} \sum_{k \in S} \gamma_i \lambda_k(i,i';p) $$
are the eigenvalues of $\kappa(\bar{\sigma}) = \sum_{p=1}^d \pi_p\, \bar{\sigma}\, \pi_p $ and $|
q_p - \tilde{q}_p| < \theta $ since $$ \| \kappa(\tilde{\sigma}) - \bar{\sigma} \|_\infty \leq \|
\tilde{\sigma} - \bar{\sigma} \|_\infty < \theta. $$  Now
\begin{eqnarray*} \lefteqn{1 - \tr \left( \sigma^{(m)} \Pi_{S,\delta + \theta} \right) =} \\ &=&
\sum_{i_1, \dots, i_m, i_{m+1} \in I} \sum_{(p_k)_{k \in S} \in \tilde{\cal T}_{S,\delta + \theta}}
\prod_{k \in S} \lambda_k(i_k,i_{k+1};p_k) \prod_{k \in S^c} \lambda_k(i_k,i_{k+1}) \\  &\leq &
\sum_{i_1, \dots, i_m, i_{m+1} \in I} \sum_{(p_k)_{k \in S} \in {\cal T}_{S,\delta}}
\prod_{k \in S} \lambda_k(i_k,i_{k+1};p_k) \prod_{k \in S^c} \lambda_k(i_k,i_{k+1}),
\end{eqnarray*} where $$ \lambda_k(i,i') = \tr \sigma_k(i,i') \mbox{ for } k \notin S, $$
and $$ {\cal T}_{S,\delta} = \left\{(p_k)_{k \in S}:\, \big| \#\{k \in S:\,p_k = p\} - |S| q_p \big| \leq |S|
\delta \right\}. $$
Introducing the Bernoulli variables $x_{k,p} = \delta_{p_k,p}$ for $k \in S$ and $p \leq d$, where
the sequences $(p_k)_{k \in S}$ are distributed according to
\begin{equation} \PP \left( (p_k)_{k \in S} \right) = \sum_{i_1, \dots, i_m, i_{m+1} \in I} \gamma_{i_1}
\prod_{k \in S}  \lambda_k(i_k,i_{k+1}; p_k)  \prod_{k \in S^c} \lambda_k(i_k,i_{k+1}),
\end{equation} we have
\begin{eqnarray*} \lefteqn{1 - \tr \left( \sigma^{(m)} \Pi_{S,\delta} \right) \leq} \\ &\leq &
\PP \left( \bigcup_{p=1}^d \left\{ (p_k)_{k \in S}:\, \left| \sum_{k \in S} x_{k,p} -
|S| q_p \right| > |S| \delta \right\} \right) \\
&\leq & \frac{1}{\delta^2} \sum_{p=1}^d \EE \left[ \left( \frac{1}{|S|} \sum_{k \in S} x_{k,p} -
q_p \right)^2 \right] \\ &=& \frac{1}{\delta^2} \sum_{p=1}^d \sum_{i_1,\dots,i_{m+1} \in I}
\sum_{(p_k)_{k \in S}} \gamma_{i_1} \prod_{k \in S}  \lambda_k(i_k,i_{k+1}; p_k)  \prod_{k \in S^c}
\lambda_k(i_k,i_{k+1})
\\ && \qquad \times \left( \frac{1}{|S|^2} \sum_{k,l \in S} x_{k,p} x_{l,p} - \frac{2}{|S|} \sum_{k
\in S} x_{k,p} q_p + q_p^2 \right) 
\end{eqnarray*}
Using the identity (\ref{consistency}),we have
$$ \sum_{i_1 \in I} \sum_{p} \gamma_{i_1} \lambda_k(i_1,i_2; p) = \gamma_{i_2}, $$
and from (\ref{sigmastate}),  $$ \sum_{i_2 \in I} \sum_p \lambda_k(i_1,i_2; p) = 1 \mbox{ and }
\sum_{i' \in I} \lambda_k(i,i') = 1.  $$ Hence we can write
\begin{eqnarray*}  && \sum_{i_1,\dots,i_{m+1} \in I}
\sum_{(p_k)_{k \in S}} \gamma_{i_1} \prod_{k \in S}  \lambda_k(i_k,i_{k+1}; p_k)  \prod_{k \in S^c}
\lambda_k(i_k,i_{k+1}) \frac{1}{|S|} \sum_{k \in S} x_{k,p} = \\ && \qquad \qquad = \frac{1}{|S|}
\sum_{k \in S} \sum_{i_k, i_{k+1} \in I}  \gamma_{i_k} \lambda_k(i_k,i_{k+1}; p) = q_p.
\end{eqnarray*}
Similarly,
\begin{eqnarray*} && \sum_{i_1,\dots,i_{m+1} \in I}
\sum_{(p_k)_{k \in S}} \gamma_{i_1} \prod_{k \in S}  \lambda_k(i_k,i_{k+1}; p_k)  \prod_{k \in S^c}
\lambda_k(i_k,i_{k+1}) \frac{1}{|S|^2} \sum_{k, l \in S;\, k < l} x_{k,p} x_{l,p} \\
&& \quad =  \frac{1}{|S|^2} \sum_{k,l \in S;\,k < l} \sum_{i_k, \dots, i_{l+1} \in I} \gamma_{i_k}
\lambda_k(i_k,i_{k+1}; p) \\ && \qquad \qquad \times \prod_{r=k+1}^{l-1} \lambda_r(i_r,i_{r+1})
\,\lambda_l(i_l,i_{l+1}; p).
\end{eqnarray*}
We now use the assumption (\ref{sigmabound}) according to which
$$ \lambda_k(i_k,i_{k+1}; p) < (1+\delta^3) \gamma_{i_{k+1}} \lambda_k(i_k; p). $$ This yields
\begin{eqnarray*}  && \frac{2}{|S|^2} \sum_{k,l \in S;\, k < l} \sum_{i_k, \dots, i_{l+1} \in I} \sum_{(p_r)_{r \in
S;\, k<r<l}} \gamma_{i_k} \lambda_k(i_k,i_{k+1}; p) \\ && \qquad \qquad \prod_{r=k+1}^{l-1} \lambda_r(i_r,i_{r+1})
\lambda_l(i_l,i_{l+1}; p) \\ &\leq & \frac{2(1+\delta^3)}{|S|^2} \sum_{k,l \in S;\, k < l}
\sum_{i_k, \dots, i_{l+1} \in I} \gamma_{i_k} \lambda_k(i_k; p) \\
&& \qquad \qquad \gamma_{i_{k+1}} \prod_{r=k+1}^{l-1} \lambda_r(i_r,i_{r+1})  \, \lambda_l(i_l,i_{l+1}; p) \\ &=& \frac{2(1+\delta^3)}{|S|^2}
\sum_{k,l \in S;\, k < l} \sum_{i_k, i_l, i_{l+1} \in I}  \gamma_{i_k} \lambda_k(i_k; p)
\gamma_{i_l} \lambda_l(i_l,i_{l+1}; p) \\ &\leq & \frac{1+\delta^3}{|S|^2} \sum_{k,l \in S}
\sum_{i_k, i_l \in I} \gamma_{i_k} \lambda_k(i_k; p) \gamma_{i_l} \lambda_l(i_l; p)  \\
&=& (1+\delta^3) q_p^2. \end{eqnarray*} Inserting, we obtain
\begin{equation} 1 - \tr \left( \sigma^{(m)} \Pi_{S,\delta} \right) =
\delta \sum_{p=1}^d q_p^2 + \frac{1}{|S| \delta^2} \sum_{p=1}^d q_p < 2 \delta \end{equation} if
$|S| > \delta^{-3}.$

\end{IEEEproof}

\begin{lemma}[Continuity]\label{lemma:Continuity}
Let $\rho,\sigma$ be states with $\|\rho-\sigma\|_1\leq\theta\leq\dfrac{1}{2}$. Then
\begin{equation}
| H(\rho)-H(\sigma) | \leq -\theta \log \frac{\theta}{d} = d \eta \left(\frac{\theta}{d}\right).
\end{equation}
\end{lemma}

\begin{lemma}[Tender operator]\label{lemma:TenderOperator}
Let $\rho$ be a state, and $X$ a positive operator with $X \leq \mathds{1}$ and $1-\tr(\rho
X)\leq\lambda\leq 1$. Then
\begin{equation}
\left\|\rho-\sqrt{X}\rho\sqrt{X}\right\|_1\leq \sqrt{8\lambda}.
\end{equation}
\end{lemma}

\begin{lemma}\label{lemma:EntropyTypical}
Let $\rho_k^{(n_0)}$  and $\eta_k^{(l_0)}$ ($k=1,\dots,m$) be states on $\calH^{\otimes n_0}$ and $\calH^{\otimes l_0}$ respectively, and $\rho$ a state on $\calH^{\otimes (n-(n_0+l_0)m)}$.
Define, for $\delta > 0$ (and $m \in \NN$), the entropy--typical projector by
\begin{equation} \Pi_{{\rm ent},\delta} = \bigoplus_{(p_1,\dots,p_m) \in
{\cal S}_\delta} (\pi_{1,p_1} \otimes \one^{(l_0)}) \otimes \dots \otimes (\pi_{m,p_m} \otimes
\one^{(l_0)}) \otimes \one^{(n-(n_0+l_0)m)},
\end{equation} where $\pi_{k,p}$ are the eigenprojections of $\Phi^{(n_0)}(\rho_k^{(n_0)})$ with
eigenvalues $\lambda_{k,p_k}$, i.e. \begin{equation} \Phi^{(n_0)}(\rho_k^{(n_0)})
= \sum_{p=1}^{d^{n_0}} \lambda_{k,p} \pi_{k,p}, \end{equation} and
\begin{equation} {\cal S}_\delta = \left\{ (p_1,\dots,p_m):\, \left| \sum_{k=1}^m \big(\log
\lambda_{k,p_k} + S(\Phi^{(n_0)}(\rho_k^{(n_0)}) \big) \right| \leq m \delta \right\}.
\end{equation} Then, for $m$ large enough,
$$ \tr \left( \Phi^{(n)}\left( \bigotimes_{k=1}^m (\rho_k^{(n_0)} \otimes \eta_k^{(l_0)}) \otimes \rho \right)
\, \Pi_{{\rm ent},\delta} \right) > 1-\delta $$ and moreover
\begin{equation}
\Pi_{{\rm ent},\delta}\,\Phi^{(n)}\left( \bigotimes_{k=1}^m (\rho_k^{(n_0)} \otimes \eta_k^{(l_0)})
\otimes \rho \right) \Pi_{{\rm ent},\delta}
         \leq \prod_{k=1}^m 2^{S(\Phi^{(n_0)}(\rho_k^{(n_0)})) + 2\delta} \,\Pi_{{\rm ent},\delta}.
         \end{equation}
\end{lemma}

\begin{IEEEproof} Analogous to (\ref{expansion}) we have the expansion
\begin{eqnarray*} \lefteqn{ \tr \left( \Phi^{(n)}\left( \bigotimes_{k=1}^m (\rho_k^{(n_0)} \otimes \eta_k^{(l_0)}) \otimes \rho \right)\,\Pi_{{\rm ent},\delta} \right) =} \\
&=& \sum_{i_1,\dots,i_{m+1} \in I} \gamma_{i_1} \tr \left( (\sigma_1^{(n_0)}(i_1,i_2) \otimes \dots
\otimes \sigma_m^{(n_0)}(i_m,i_{m+1}))\, \tilde{\Pi}_{{\rm ent},\delta} \right) \end{eqnarray*} where
$$ \tilde{\Pi}_{{\rm ent},\delta} = \bigoplus_{(p_1,\dots,p_m) \in {\cal S}_\delta} \pi_{p_1} \otimes \dots
\otimes \pi_{p_m}, $$ and $\sigma_k^{(n_0)}(i,i')$ are given by
\begin{eqnarray*} \sigma_k^{(n_0)}(i,i')  &=& \sum_{i_2,\dots,i_{n_0}}  q_{i,i_2} q_{i_2 i_3} \dots
q_{i_{n_0-1},i_{n_0}} q^{(l_0)}_{i_{n_0},i'}
\\ && \qquad \times (\Phi_{i} \otimes \dots \otimes \Phi_{i_{n_0}})(\rho_k^{(n_0}).
\end{eqnarray*}


As in the proof of Lemma~\ref{lemma:MarkovWeakLaw} we apply the map $\kappa^{\otimes m}$ and obtain
\begin{eqnarray*} \lefteqn{ \tr \left( \Phi^{(n)}(f^{(n)}(w))\,\Pi_{S,\delta} \right) =} \\
&=& \sum_{i_1,\dots,i_{m+1} \in I} \gamma_{i_1} \sum_{(p_1,\dots,p_m) \in {\cal S}_\delta}
\prod_{k=1}^m \tr \left(  \pi_{k,p_k} \sigma_k^{(n_0)}(i_k,i_{k+1})  \pi_{k,p_k} \right).
\end{eqnarray*}
Introducing the probability distribution on $\{1,\dots,d^{n_0}\}^m$ given by
\begin{equation} \PP (A) = \sum_{i_1,\dots,i_{m+1} \in I} \gamma_{i_1} \sum_{(p_1,\dots,p_m) \in A}
\prod_{k=1}^m \tr \left(  \pi_{k,p_k} \sigma_k^{(n_0)}(i_k,i_{k+1})  \pi_{k,p_k} \right)
\end{equation} for $A \subset \{1,\dots,d^{n_0}\}^m$, we have
\begin{eqnarray*} \lefteqn{ \tr \left( \Phi^{(n)}(f^{(n)}(w))\,\Pi_{S,\delta} \right) = \PP({\cal
S}_\delta)=} \\ &=& \PP \left( \left| \frac{1}{m} \sum_{k=1}^m \left( \log \lambda_{k,p_k} +
S(\Phi^{(n_0)}(\rho_{k,1}(w) \otimes \dots \otimes \rho_{k,n_0})) \right) \right| \leq \delta
\right).
\end{eqnarray*}
As before the sum in the expectation of $\log \lambda_{k,p_k}$ telescopes, and \begin{equation}
\begin{split}
\EE (\log \lambda_{k,p_k}) &= \sum_{i_k,i_{k+1} \in I} \gamma_{i_k} \sum_{p=1}^{d^{n_0}} \tr \left(
\pi_{k,p} \sigma_k^{(n_0)}(i_k,i_{k+1}) \right) \log \lambda_{k,p} \\ &= \sum_{p=1}^{d^{n_0}} \tr
\left( \pi_{k,p} \Phi^{(n_0)}(\rho_k^{(n_0)}) \right) \log
\lambda_{k,p} \\ & = - S(\Phi^{(n_0)}(\rho_k^{(n_0)})).
\end{split}
\end{equation}
To show that $\PP({\cal S}_\delta) \to 0$, we compute the variance of $\frac{1}{m} \sum_{k=1}^m
\log \lambda_{k,p_k}$ as before. We have
\begin{equation} \begin{split} & \EE \left( \left| \frac{1}{m} \sum_{k=1}^m \left( \log \lambda_{k,p_k} +
S(\Phi^{(n_0)}(\rho_{k,1}(w) \otimes \dots \otimes \rho_{k,n_0})) \right) \right|^2 \right) = \\
&= \frac{1}{m^2} \sum_{k,l=1}^m \EE \left( \log \lambda_{k,p_k}\, \log \lambda_{l,p_l} \right) -
\left(\frac{1}{m} \sum_{k=1}^m S(\Phi^{(n_0)}(\rho_k^{(n_0)}))
\right)^2. \end{split} \end{equation}
The term $k=l$ yields
$$ \frac{1}{m^2} \sum_{k=1}^m \sum_{p=1}^{d^{n_0}} \lambda_{k,p} (\log \lambda_{k,p})^2 \to 0 $$
since the sum $\sum_{p=1}^{d^{n_0}} \lambda_{k,p} (\log \lambda_{k,p})^2$ is bounded by $n_0^2
(\log d)^2$. For the terms $k < l$ we use the argument of Lemma~\ref{lemma:MarkovWeakLaw}: By the
assumption (\ref{sigmabound})
$$ \lambda_k(i_k,i_{k+1}; p) < (1+\delta^3) \gamma_{i_{k+1}} \lambda_k(i_k; p). $$ Writing $\lambda_k(i,i';p) =
\tr( \pi_{k,p} \sigma_k(i,i'))$, this yields
\begin{equation} \begin{split}
&\frac{2}{m^2} \sum_{1 \leq k < l \leq m} \EE \left( \log \lambda_{k,p_k}\, \log \lambda_{l,p_l}
\right) = \\ &= \frac{2}{m^2} \sum_{1 \leq k < l \leq m} \sum_{i_k, \dots, i_{l+1} \in I}
\sum_{p_k, \dots, p_l = 1}^{d^{n_0}} \gamma_{i_k} \prod_{r=k}^l \lambda_r(i_r,i_{r+1}; p_r)
\log \lambda_{k,p_k}\, \log \lambda_{l,p_l} \\
 &\leq  \frac{2(1+\delta^3)}{m^2} \sum_{1 \leq k < l \leq m}
\sum_{i_k, \dots, i_{l+1} \in I} \sum_{p_k, \dots, p_l = 1}^{d^{n_0}} \gamma_{i_k} \lambda_k(i_k; p_k) \log \lambda_{k,p_k}
\gamma_{i_{k+1}} \prod_{r=k+1}^l \lambda_r(i_r,i_{r+1}; p_r)  \log \lambda_{l,p_l} \\
&= \frac{2(1+\delta^3)}{m^2} \sum_{1 \leq k < l \leq m}
 \sum_{p_k = 1}^{d^{n_0}}  \lambda_{k, p_k}\, \log \lambda_{k,p_k} \sum_{i_l, i_{l+1} \in I} \sum_{p_l = 1}^{d^{n_0}} \gamma_{i_{l}} \lambda_l(i_l,i_{l+1}; p_l)  \log \lambda_{l,p_l}
\\ &= \frac{2(1+\delta^3)}{m^2} \sum_{1 \leq k < l \leq m}
S(\Phi^{(n_0)}(\rho_k^{(n_0)})) \,S(\Phi^{(n_0)}(\rho_l^{(n_0)}))  \\
&\leq (1+\delta^3) \left( \sum_{k=1}^m S(\Phi^{(n_0)}(\rho_k^{(n_0)})) \right)^2. \end{split} \end{equation}

To prove the second bound, we write
\begin{eqnarray*} \lefteqn{ \Phi^{(n)}\left( \bigotimes_{k=1}^m (\rho_k^{(n_0)} \otimes \eta_k^{(l_0)}) \otimes \rho \right) =}
\\ &=& \sum_{i_1,\dots,i_{m+1} \atop i'_1,\dots,i'_{m+1}} \gamma_{i_1} \sigma_1^{(n_0)}(i_1,i'_1) \otimes
\sigma^{(l_0)}_1(i'_1,i_2) \otimes \dots \\ && \qquad \otimes  \sigma_m^{(n_0)}(i_m,i'_m) \otimes
\sigma^{(l_0)}_1(i'_m,i_{m+1}) \otimes \sigma^{(l)}(i_{m+1},i'_{m+1}),
\end{eqnarray*}
where \textit{now}
\begin{equation} \sigma^{(n_0)}_k(i,i') = \sum_{i_2,\dots,i_{n_0}} q_{ii_1} \dots q_{i_{n_0},i'} (\Phi_{i} \otimes \dots \otimes \Phi_{i_{n_0}})(\rho_k^{(n_0)})
\end{equation}
and
\begin{equation} \sigma^{(l_0)}_k(i,i') = \sum_{i_2,\dots,i_{l_0}} q_{ii_1} \dots q_{i_{l_0},i'} (\Phi_{i} \otimes \dots \otimes \Phi_{i_{l_0}})(\eta_k^{(l_0)})
\end{equation}
and with $l = n-m(n_0+l_0)$,
\begin{equation} \sigma^{(l)}(i,i') = \sum_{i_2,\dots,i_{l}} q_{ii_1} \dots q_{i_{l},i'} (\Phi_{i} \otimes \dots \otimes \Phi_{i_{l}})(\rho).
\end{equation}
Using the inequality
\begin{equation}
\begin{split}
& \Pi_{{\rm ent}, \delta} \, \bigotimes_{k=1}^m \left( \Phi^{(n_0)}(\rho_k^{(n_0)}) \otimes \one^{(l_0)})
\right) \otimes \one^{(l)} \, \Pi_{{\rm ent}, \delta} \leq \\
&\leq \prod_{k=1}^m 2^{- S(\Phi^{(n_0)}(\rho_k^{(n_0)})) + \delta} \, \Pi_{{\rm ent}, \delta}, \end{split} \end{equation}
which follows from the definition of $\Pi_{{\rm ent},\delta}$, together with the bound
$$ \sigma_k^{(n_0)}(i,i') \leq (1+\delta^3) \gamma_{i'} \sum_{i'' \in I} \sigma_k^{(n_0)}(i,i''), $$
we find

\begin{equation}
\begin{split}
& \Pi_{{\rm ent},\delta}\,\Phi^{(n)}\left( \bigotimes_{k=1}^m (\rho_k^{(n_0)} \otimes \eta_k^{(l_0)}) \otimes \rho \right) \Pi_{{\rm ent},\delta} \\
         & \qquad \leq (1+\delta^3)^m
\Pi_{{\rm ent}, \delta} \, \bigotimes_{k=1}^m \left( \Phi^{(n_0)}(\rho_k^{(n_0)}) \otimes \one^{(l_0)})
\right) \otimes \one^{(l)} \, \Pi_{{\rm ent}, \delta} \leq \\
&\leq \prod_{k=1}^m 2^{- S(\Phi^{(n_0)}(\rho_k^{(n_0)})) + 2 \delta} \, \Pi_{{\rm ent}, \delta}, \end{split} \end{equation}

\end{IEEEproof}

\begin{lemma}[Shadow bound] \label{lemma:ShadowBound}
Suppose that $0 \leq \Lambda \leq \unity$ and $\rho$ is a state such that for constants $\lambda,
\mu_1, \mu_2 > 0$,
$$ \tr(\rho \Lambda) > 1-\lambda \mbox{ and } \mu_1 \Lambda \leq \Lambda^{1/2} \rho \lambda^{1/
2} \leq \mu_2 \Lambda, $$ then
$$ (1-\lambda) \mu_2^{-1} \leq \tr \Lambda \leq \mu_1^{-1} $$
and for $B \geq 0$,
$$ \tr (\rho B) \geq \eta \implies \tr B \geq (\eta - \sqrt{8 \lambda})
\mu_2^{-1}. $$
\end{lemma}

\end{document}